\documentclass[11pt,a4paper]{article}

\usepackage{jheppub}
\usepackage{amsmath}
\usepackage{amsfonts}
\usepackage{amssymb}
\usepackage{tikz}
\usetikzlibrary{arrows,shapes}

\usepackage[compat=1.1.0]{tikz-feynman}

\usepackage{amssymb}
\usepackage{color}

\allowdisplaybreaks
\newcommand\numberthis{\addtocounter{equation}{1}\tag{\theequation}}

\usepackage[utf8]{inputenc}
\DeclareUnicodeCharacter{27F9}{\ensuremath{\Longrightarrow}}
\DeclareUnicodeCharacter{2209}{\ensuremath{\notin}}
\DeclareUnicodeCharacter{266D}{\ensuremath{\flat}}
\DeclareUnicodeCharacter{2208}{\ensuremath{\in}}
\DeclareUnicodeCharacter{00E9}{\'{e}}
\DeclareUnicodeCharacter{2264}{\ensuremath{\leq}}
\DeclareUnicodeCharacter{2265}{\ensuremath{\geq}}
\DeclareUnicodeCharacter{2200}{\ensuremath{\forall}}
\DeclareUnicodeCharacter{221E}{\ensuremath{\infty}}
\DeclareUnicodeCharacter{221D}{\ensuremath{\propto}}
\DeclareUnicodeCharacter{210F}{\ensuremath{\hbar}}
\DeclareUnicodeCharacter{2020}{\ensuremath{\dagger}}
\DeclareUnicodeCharacter{03B1}{\ensuremath{\alpha}}
\DeclareUnicodeCharacter{03B2}{\ensuremath{\beta}}
\DeclareUnicodeCharacter{0393}{\ensuremath{\Gamma}}
\DeclareUnicodeCharacter{03B3}{\ensuremath{\gamma}}
\DeclareUnicodeCharacter{03B8}{\ensuremath{\theta}}
\DeclareUnicodeCharacter{21D2}{\ensuremath{\Rightarrow}}
\DeclareUnicodeCharacter{0394}{\ensuremath{\Delta}}
\DeclareUnicodeCharacter{03B4}{\ensuremath{\delta}}
\DeclareUnicodeCharacter{03C1}{\ensuremath{\rho}}
\DeclareUnicodeCharacter{03BE}{\ensuremath{\xi}}
\DeclareUnicodeCharacter{03C0}{\ensuremath{\pi}}
\DeclareUnicodeCharacter{03A0}{\ensuremath{\Pi}}
\DeclareUnicodeCharacter{03BC}{\ensuremath{\mu}}
\DeclareUnicodeCharacter{03BD}{\ensuremath{\nu}}
\DeclareUnicodeCharacter{222B}{\ensuremath{\int}}
\DeclareUnicodeCharacter{2211}{\ensuremath{\Sigma}}
\DeclareUnicodeCharacter{03C3}{\ensuremath{\sigma}}
\DeclareUnicodeCharacter{03C4}{\ensuremath{\tau}}
\DeclareUnicodeCharacter{03BB}{\ensuremath{\lambda}}
\DeclareUnicodeCharacter{03B7}{\ensuremath{\eta}}
\DeclareUnicodeCharacter{03A6}{\ensuremath{\Phi}}
\DeclareUnicodeCharacter{03D5}{\ensuremath{\phi}}
\DeclareUnicodeCharacter{03A8}{\ensuremath{\Psi}}
\DeclareUnicodeCharacter{03C8}{\ensuremath{\psi}}
\DeclareUnicodeCharacter{03F5}{\ensuremath{\epsilon}}
\DeclareUnicodeCharacter{2202}{\ensuremath{\partial}}
\DeclareUnicodeCharacter{220F}{\ensuremath{\prod}}
\DeclareUnicodeCharacter{03C9}{\ensuremath{\omega}}
\DeclareUnicodeCharacter{2260}{\ensuremath{\neq}}
\DeclareUnicodeCharacter{22C5}{\ensuremath{\cdot}}
\DeclareUnicodeCharacter{27E8}{\ensuremath{\langle}}
\DeclareUnicodeCharacter{27E9}{\ensuremath{\rangle}}
\DeclareUnicodeCharacter{2192}{\ensuremath{\rightarrow}}
\DeclareUnicodeCharacter{00B2}{\ensuremath{{}^2}}
\DeclareUnicodeCharacter{221A}{\ensuremath{\sqrt{}}}
\DeclareUnicodeCharacter{223C}{\ensuremath{\sim}}
\DeclareUnicodeCharacter{226A}{\ensuremath{\ll}}
\DeclareUnicodeCharacter{00D7}{\ensuremath{\times}}

\begin{document}

\preprint{FR-PHENO-2020-017, IPPP/20/50}

\title{Two-Loop Five-Parton Leading-Colour Finite Remainders in the Spinor-Helicity Formalism}

\author[a]{Giuseppe De Laurentis,}
\emailAdd{giuseppe.de.laurentis@physik.uni-freiburg.de}
\affiliation[a]{Physikalisches Institut, Albert-Ludwigs-Universität at Freiburg, Hermann-Herder.Str. 3, D-79104 Freiburg, Germany}

\author[b]{Daniel Ma\^{\i}tre}
\emailAdd{daniel.maitre@durham.ac.uk}
\affiliation[b]{Department of Physics, University of Durham, Durham DH1 3LE, UK}

\date{\today}

\abstract{
  We present all two-loop five-parton leading-colour finite remainders in the spinor-helicity formalism by analysing numerical evaluations of their known expressions in terms of Mandelstam invariants. Recasting them in terms of spinor-helicity variables allows us to obtain expressions which are more compact, faster to evaluate, numerically more stable and manifestly free from poles of higher order than necessary. At the same time, due to the better scaling of our reconstruction strategy with the complexity of the input, we required one order of magnitude fewer numerical samples to complete the analytical reconstruction than were needed by the authors of Ref.~\cite{Abreu:2019odu}, albeit using higher numerical working precision. This places our reconstruction technique as an alternative to the finite-field single-numerator reconstruction for future applications.
}


\maketitle

\section{Introduction}

Scattering amplitudes are the heart of any perturbative prediction for collider physics. The increase in luminosity and precision on the experimental side is driving the theoretical effort to match such results. This means calculations are being pushed towards higher loop order and higher multiplicities. While there exist many methods to compute scattering amplitudes that work in principle, practical implementations often run into complexity and performance bottlenecks. More efficient alternative methods to calculate matrix elements are an important driver of practical progress in the quest for precision predictions, as they bring more complex calculations within operational reach.

In recent years, significant progress has been made in the computation of two-loop amplitudes with five massless external legs \cite{Abreu:2019odu, Abreu:2020xvt, Badger:2015lda, Gehrmann:2015bfy, Dunbar:2016aux, Badger:2017jhb, Abreu:2017hqn, Abreu:2018jgq, Badger:2018enw, Abreu:2018zmy, Chawdhry:2018awn}, especially by means of generalised unitarity methods \cite{Bern:1994zx,Bern:1994cg,Bern:1997sc,Britto:2004nc,Ossola:2006us,Ellis:2007br,Giele:2008ve,Berger:2008sj,Ita:2015tya,Abreu:2017idw}. Calculations such as these are performed numerically to circumvent practical bottlenecks in intermediate stages of the analytical calculation. Despite the success of these methods, numerical routines can still require up to several minutes to be executed at every given phase-space point. For any phenomenological application to be feasible, it is crucial to find a more efficient evaluation path. This is done by reconstructing reasonably compact analytical expression from a finite number of numerical evaluations. 

Analytical expressions for the planar two-loop five-parton amplitudes were obtained in Ref.~\cite{Abreu:2019odu} by means of a modified Newton method over finite-fields \cite{Peraro:2016wsq}. We show that by analysing the singularity structure of the coefficients and performing the reconstruction in terms of spinor-helicity variables as proposed in Ref.~\cite{DeLaurentis:2019phz} the obtained analytical expressions are more compact, faster to evaluate and numerically more stable. This is achieved while reducing the number of required numerical samples by about an order of magnitude.

In the following sections, we briefly review the original calculation to introduce the specific quantities under consideration, then review the spinor-helicity reconstruction strategy, and analyse the algebraic complexity of both reconstruction and final results. Lastly, we consider the numerical stability of the results for phenomenological applications. 

\subsection{Summary of the original calculation}\label{sec:summary-of-calculation}

The original calculation in Ref.~\cite{Abreu:2019odu} was performed via numerical $D$-dimensional generalised unitarity in the leading-colour approximation, more precisely in the limit of large number of colours $N_c$ while keeping the ratio to the number of flavours $N_f$ fixed.

The full amplitudes $A$ are stripped of their dependence on the gauge-group factors by colour-ordering decompositions \cite{Ochirov:2019mtf, Ochirov:2016ewn} in terms of color-ordered sub-amplitudes $\mathcal{A}$. Their expansion in terms of the bare QCD coupling  $α_0$ reads
\begin{equation}
  \mathcal{A} = g_0^3 \Big( \mathcal{A}^{(0)} + \frac{α_0}{4π} N_c \mathcal{A}^{(1)} + \left(\frac{α_0}{4π}\right)^2 N_c^2 \mathcal{A}^{(2)} + \mathcal{O}(α_0^3) \Big) \, ,
\end{equation}
where $\mathcal{A}^{(n)}$ is a $n$-loop amplitude. Each loop amplitude can be further expanded in the ratio $N_f/N_c$ as
\begin{eqnarray}
  \mathcal{A}^{(1)}&=&\mathcal{A}^{(1)[N_f^0]} + \frac{N_f}{N_c}\mathcal{A}^{(1)[N_f^1]} \, , \label{eq:nf_expansion_1l} \\
  \mathcal{A}^{(2)}&=&\mathcal{A}^{(2)[N_f^0]} + \frac{N_f}{N_c}\mathcal{A}^{(2)[N_f^1]} + \left(\frac{N_f}{N_c}\right)^2\mathcal{A}^{(2)[N_f^2]} \, . \label{eq:nf_expansion_2l}
\end{eqnarray}

Bare quantities are related to renormalised ones through the running of the QCD coupling $α_s(μ)$, whose relation to the bare coupling $α_0$ is given by the $β$-function 
\begin{equation}
  α_0μ_0^{2ϵ}S_{ϵ} =α_sμ^{2ϵ}\left( 1-\frac{β_0}{ϵ}\frac{α_s}{4π} + \left(\frac{β_0^2}{ϵ^2}-\frac{β_1}{ϵ} \right)  \left( \frac{α_s}{4π}\right )^2 + \mathcal{O} \left( α_s^3 \right) \right).
\end{equation}
with $S_ϵ=(4π)^{ϵ}e^{-ϵγ_E}\,$. This, in turn, leads to the renormalised amplitude
\begin{equation}
  \mathcal{A}_R = S_ϵ^{-\frac{3}{2}} g_s^3\left( \mathcal{A}_R^{(0)} +\frac{α_s}{4π} N_c \, \mathcal{A}_R^{(1)} +\left(\frac{α_s}{4π}\right)^2 N_c^2\mathcal{A}_R^{(2)} +\mathcal{O}(α_s^3) \right) \, .
\end{equation}

The $ϵ$-pole dependence of an $n$-loop amplitude is determined by lower-loop amplitudes and well known universal factors \cite{Catani:1998bh, Sterman:2002qn, Becher:2009cu, Gardi:2009qi}
\begin{eqnarray}
    \mathcal{A}_R^{(1)}&=&{\bf I}^{(1)}_{[n]}(\epsilon) \mathcal{A}_R^{(0)}+\mathcal{O} (\epsilon^0)\,,\\
    \mathcal{A}_R^{(2)}&=&{\bf I}^{(2)}_{[n]}(\epsilon) \mathcal{A}_R^{(0)}+{\bf I}^{(1)}_{[n]}(\epsilon) \mathcal{A}_R^{(1)}+\mathcal{O}(\epsilon^0)\,. \label{eq:two-loop-poles}
\end{eqnarray}
The latter equation can be rearranged to obtain the definition of the so-called finite remainders, which contain the genuine two-loop information
\begin{equation}\label{eq:finite_remainder_2l}
  \mathcal{R}^{(2)}=\mathcal{A}_R^{(2)} -{\bf I}_{[n]}^{(1)}\mathcal{A}_R^{(1)} -{\bf I}_{[n]}^{(2)}\mathcal{A}_R^{(0)} + \mathcal{O}(\epsilon)\,.
\end{equation}

These finite remainders can be expressed as sum of products of rational coefficients $r_i$ and special transcendental functions $h_i$ such as pentagon functions \cite{Gehrmann:2018yef, Chicherin:2020oor}
\begin{equation}\label{eq:sum_over_coeff_pentagon_func}
  \mathcal{R}^{(2)} = \sum_{i} r_i h_i \,.
\end{equation}
Eq.~\ref{eq:finite_remainder_2l} and \ref{eq:sum_over_coeff_pentagon_func} are valid term by term in the $N_f/N_c$ expansion of Eq.~\ref{eq:nf_expansion_2l}. The rational coefficients $r_i$ are the subject of the present study.

\section{Spinor-Helicity Remainders}

We make use of the strategies introduced in Ref.~\cite{DeLaurentis:2019phz} to construct spinorial expressions for all 740 function coefficients of the two-loop five-parton finite remainders (see Eq.~\ref{eq:sum_over_coeff_pentagon_func}). These are labelled by particle content, helicity, position in the $N_f/N_c$ expansion, and by a numerical index. For simplicity's sake, we will refer to them by single generic index $i$. Explicit examples of what is achievable with this type of reconstruction are given in Section~\ref{sec:examples} of the Appendix. The full results can be found in the ancillary files, which can be imported into \verb!Mathematica! and evaluated with \verb!S@M! \cite{Maitre:2007jq}.

\subsection{Reconstruction strategy}

The finite-field reconstruction originally employed in Ref.~\cite{Abreu:2019odu} requires to choose a minimal set of linearly independent invariants. This minimal set was either momentum twistor parameters \cite{Hodges:2009hk}, or a combination of Mandelstam variables, $s_{ij}=2P_i\cdot P_j$, and Gram determinants, $\text{tr}_5(ijkl)=\text{tr}(γ_5P_iP_jP_kP_l)$. In the case of our input expressions the choice for the five-point amplitudes was
\begin{gather}
  \{ s_{12}, s_{23}, s_{34}, s_{45}, s_{15}, \text{tr}_5(1234) \} .
\end{gather}
To uncover the simpler expressions, we relax this condition and instead consider an over-complete (in a linear sense) set of spinor variables\footnote{For an introduction to the spinor-helicity notation please see Ref.~\cite{Dixon:2013uaa, Maitre:2007jq}.}, which for the current application can be taken to be
\begin{equation}\label{eq:list_of_spinor_variables}
  \begin{array}{c@{\,}c@{\,}c}
  \vec{v} = \{   & ⟨12⟩, ⟨13⟩, ⟨14⟩, ⟨15⟩,⟨23⟩, ⟨24⟩, ⟨25⟩, ⟨34⟩, ⟨35⟩, ⟨45⟩, & \\
                 & [12], [13], [14], [15], [23], [24], [25], [34], [35], [45], & \\
                 & ⟨1|2+3|1], ⟨1|2+5|1],  ⟨2|1+3|2], ⟨2|1+5|2], ⟨3|1+2|3], & \\
                 & ⟨3|1+5|3], ⟨4|1+2|4], ⟨4|1+5|4], ⟨5|1+2|5], ⟨5|1+4|5]\phantom{,} & \} \, .
  \end{array}
\end{equation}
This set of variables was discovered by starting off with an even larger one containing all spinor contractions of a certain form and restricting it to those for which the rational functions under consideration actually display diverging behaviour.

The advantage of such a choice of variables is that there is now a one-to-one correspondence between variables and possible poles of the rational coefficients in complex momentum space. This relation can be exploited numerically, for instance via high-precision floating-point evaluations in singular limits
\begin{equation}\label{eq:singular_limits}
  \lim _{v_j \to ϵ \ll 1} r_i \propto ϵ^{α_{i,j}} \, ,
\end{equation}
where $α_{i,j}$ is the order of the pole ($α_{i,j}<0$) or zero ($α_{i,j}>0$) of the coefficient $r_i$ in the spinor variable $v_j$. If $α_{i,j}$ is zero, then $v_j$ is neither a pole nor a zero of $r_i$. 

After evaluating $r_i$ in all singular limits of Eq.~\ref{eq:singular_limits}, we obtain the least common denominator ($\mathcal{LCD}$) representation of the rational function $r_i$
\begin{equation}\label{eq:lcd}
  r_i=  \frac{\mathcal{N_{LCD}}}{\mathcal{D_{LCD}}} = \mathcal{N} \, \prod_j v^{α_{i,j}}_j \, ,
\end{equation}
where the quotient $\mathcal{N_{LCD}}/\mathcal{N}$ is given by the product of common factors in the numerator $\{v_j:α_{i,j}>0\}$ or $1$ if no common factor was found. In general, $\mathcal{N}$ is guaranteed to be free from denominator factors, assuming $\vec{v}$ contains all possible poles of $r_i$.

Note that in some cases such as the first example given in the Appendix (\ref{eq:full_example1_mandelstam}/\ref{eq:full_example1_spinor}) the task of determining the coefficient $r_i$ is rendered entirely trivial by this procedure, as the product of the $v^{α_{i,j}}_j$ factors yields the full answer up to an easily obtained constant pre-factor. More generally, the numerator $\mathcal{N}$ is a polynomial in spinor variables of Eq.~\ref{eq:list_of_spinor_variables} whose complexity depends on its mass dimension and its phase weights (see Sec.~2.3 of Ref.~\cite{DeLaurentis:2019phz}). The coefficients of this polynomial can be fixed by linear solving.

Let us consider the second example given in the Appendix (\ref{eq:full_example2_mandelstam}/\ref{eq:full_example2_spinor}), Eq.~\ref{eq:lcd} reads\footnote{This coefficient is for $\mathcal{R}^{(2)[N_f^1]}_{g^{-}g^{+}g^{-}g^{+}g^{+}}$.}
\begin{equation}
  r_{10} = \frac{\mathcal{N}}{[13]^4[25]^4⟨5|1+2|5]^3} \,
\end{equation}
where $\mathcal{N}$ has mass dimension $14$ and phase weights $[-4, -4, -4, 0, -4]$. In this case, the corresponding ansatz, i.e.~a set of products of spinor brackets which spans $\mathcal{N}$, only has 160 entries. However, note that this is a fairly easy example. In general, as the number of factors in $\mathcal{D_{LCD}}$ increases, so does the size of the system corresponding to $\mathcal{N}$, and in some cases it can exceed $100\,000$ entries.

In order to simplify the reconstruction, we need to go one step further and perform a partial-fraction decomposition before actually attempting the numerator reconstruction. Such a decomposition will also better represent the pole structure of the coefficients. Thus, let us write the coefficients in the following form
\begin{equation}\label{eq:partial_fractions}
  r_i= \sum_k \mathcal{N}_k \, \prod_{j} v_j^{β^{k}_{i,j}} \quad \text{with} \quad β^{k}_{i,j} ≥ α_{i,j} \, .
\end{equation}
This decomposition is of course not unique. If we denote by $\big[\mathcal{N}\big]$ the size of the ansatz for the expression $\mathcal{N}$, the aim is to have $\sum_k \big[\mathcal{N}_k\big] \ll \big[\mathcal{N}\big]$. This is generally achieved by picking the $β^k_{i,j}$ as large as possible.

In order to obtain insights into the structure of possible partial-fraction decomposition, we study the behaviour of $r_i$ in all doubly singular limits
\begin{equation}\label{eq:doubly_singular_limits}
  \lim _{v_{j_1}, v_{j_2} \to ϵ \ll 1} r_i = ϵ^{α_{i,(j_1,j_2)}} \, .
\end{equation}
By comparing $α_{i,(j_1, j_2)}$ to $α_{i,j_1}+α_{i,j_2}$ we can infer potential partial-fraction decompositions. Since in general there will exist additional factors $v_{j_3≠j_1,j_2}$ that will also be of order $\sim O(ϵ)$ in this double limit, the logic to apply to extract useful information out of the table of exponents $α_{i,(j_1,j_2)}$ can be fairly lengthy. We refer the reader to Ref.~\cite{DeLaurentis:2019phz} for a detailed discussion. Let us consider an example instead.

In the case we already considered (\ref{eq:full_example2_mandelstam}/\ref{eq:full_example2_spinor}), two of the doubly singular limits are
\begin{equation}
  \lim _{[13], ⟨5|1+2|5] \to ϵ \ll 1} r = ϵ^{-7} \quad  \text{and} \quad \lim _{[25], ⟨5|1+2|5] \to ϵ \ll 1} r = ϵ^{-4} \, .
\end{equation}
Therefore, it is logical to conjecture that if we want to reconstruct the residue of $⟨5|1+2|5]^3$ we need to include in the denominator the full fourth order pole $[13]^4$ but $[25]$ may suffice as a simple pole. Indeed this is the case, as shown by the first fraction in \ref{eq:full_example2_spinor}.

This strategy appears to scale well with the size of $\mathcal{D_{LCD}}$, severely limiting $[\mathcal{N}_k]$ even in the most complicated cases. Furthermore, as explained in  Ref.~\cite{DeLaurentis:2019phz}, specific numerators $\mathcal{N}_k$ can be isolated and reconstructed individually by generating phase-space points in limits where they are dominant. This allows to solve a few small systems instead of a single large one, and to apply the strategy iteratively by recursively subtracting reconstructed partial-fractioned terms and repeating the study of the singular limits for the remainder.

\section{Results}

In this section we compare three aspects of our results compared to those we used as an input. First, we compare their complexity in terms of the leaf count of the finite remainder expressions, that is in terms of the number of nodes in their abstract syntax trees. Second, we estimate the number of calls to the numerical amplitude evaluation procedure needed by both methods for the analytical reconstruction of the rational functions. Third, we analyse the numerical stability of the resulting expressions under realistic conditions.  
\subsection{Algebraic complexity of the result}
We compare the complexity of our expressions with the original input expressions from Ref.~\cite{Abreu:2019odu} which were simplified by means of Leinartas partial fractioning \cite{leinartas1978factorization, raichev2012leinartass}. As a measure of complexity we use the leaf count.
\begin{figure}[t]
  \centering
  \includegraphics[height=67.5mm, width=120mm]{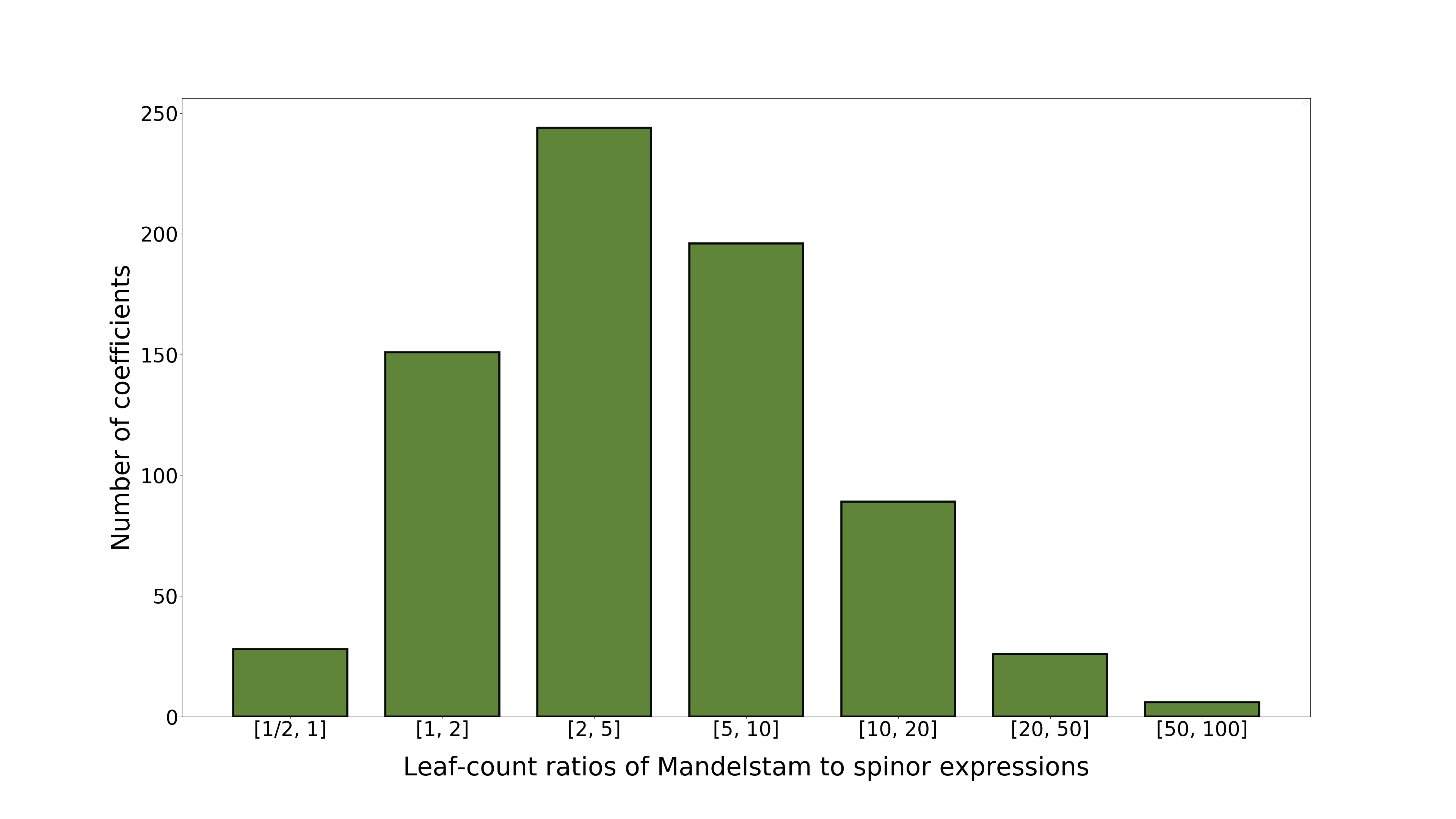}
  \caption{Ratios of leaf counts of Mandelstam expressions to spinor-helicity ones.}
  \label{figure:LeafCountRatios}
\end{figure}

Figure~\ref{figure:LeafCountRatios} shows the ratio of the leaf counts for all 740 coefficients. This comparison is between the raw output of the spinor reconstruction to the Leinartas simplified version of the Mandelstam expressions. The vast majority of the coefficients is simplified in the spinor version, in some cases by more than one order of magnitude. Overall, the total new leaf count is about two times smaller, at $4 \cdot 10^{6}$ compared to $9 \cdot 10^{6}$. This is the case, despite the leaf count of most expressions is reduced by more than a factor of two, because the largest reduction occurred for coefficients of relative small size. As a result of the reduced complexity, the total speed up for the evaluation of all coefficients is also a factor of about 2, at about $1s$ instead of $2s$ on \verb!Mathematica!.  % 4176128 vs 9162373

\subsection{Number of numerical evaluations}

For the results we present in this paper, we take as input numerical evaluations of analytical expressions previously obtained in Ref.~\cite{Abreu:2019odu}. In principle our method could have been applied directly to the numerical program from which the expressions we take as input were themselves obtained. Of course this would have required a significantly larger amount of computing resources. In this section we quantify how many numerical evaluations are needed by each method to recover the analytical expressions, and show that using our reconstruction technique would have required far fewer, albeit in some cases with higher numerical precision.

The number of evaluations required to perform the reconstruction of a numerator $\mathcal{N}_k$ is directly related to the size of the linear system which parametrises it. In particular, the relevant systems are Lorentz invariant polynomials with $[\mathcal{N}_k]$ linearly independent monomials. Let us perform the counting for a numerator of mass dimension $d$ and zero phase weights, since this case can be equivalently expressed in terms of spinor products or Mandelstam variables, $s_{ij}$, and Gram determinants, $\text{tr}_5(ijkl)$\footnote{Note that there is only one linearly independent $\text{tr}_5$ at five-point.}. Let $m$ be the multiplicity of the phase space, then the counting for the number of linearly independent $s_{ij}$ and $\text{tr}_5(ijkl)$ is simply
\begin{equation}
  [s_{ij}] = \frac{m(m-3)}{2} \quad \text{and} \quad  [tr_5(ijkl)] = {m-1 \choose 4} \, .
\end{equation}
The size of the numerator ansatz is then bounded as follows
\begin{equation}\label{eq:ansatz_size}
  \left(\mkern -9mu \begin{pmatrix}\, [s_{ij}] \, \\ \, d/2 \, \end{pmatrix} \mkern -9mu \right) \leq [\mathcal{N}_k] \leq \left(\mkern -9mu  \begin{pmatrix} \, [s_{ij}] \, \\ \, d/2 \, \end{pmatrix} \mkern -9mu \right) + [tr_5(ijkl)] \left(\mkern -9mu  \begin{pmatrix} \, [s_{ij}] \, \\ \, (d-4)/2 \, \end{pmatrix} \mkern -9mu \right) \, ,
\end{equation}
where the double parenthesis denotes combinations with replacement
\begin{equation}
  C^R(n,r) = \left(\mkern -9mu \begin{pmatrix}\, n \, \\ \, r \, \end{pmatrix} \mkern -9mu \right) = \dfrac{(n + r - 1)!}{ r! (n - 1)! } \, .
\end{equation}
Unfortunately, the exact counting valid at all-multiplicity is not easy to obtain due to the following identity
\begin{equation}
  \text{tr}_5(2345)P_1^μ-\text{tr}_5(1345)P_2^μ+\text{tr}_5(1245)P_3^μ-\text{tr}_5(1235)P_4^μ+\text{tr}_5(1234)P_5^μ=0 \; .
\end{equation}
However, the upper bound of Eq.~\ref{eq:ansatz_size} is saturated at all multiplicities for mass dimensions $d≤4$, and at all mass dimensions for multiplicities $m≤5$, which includes case relevant for the current study.

\begin{figure}[t]
  \centering
  \includegraphics[height=67.5mm, width=120mm]{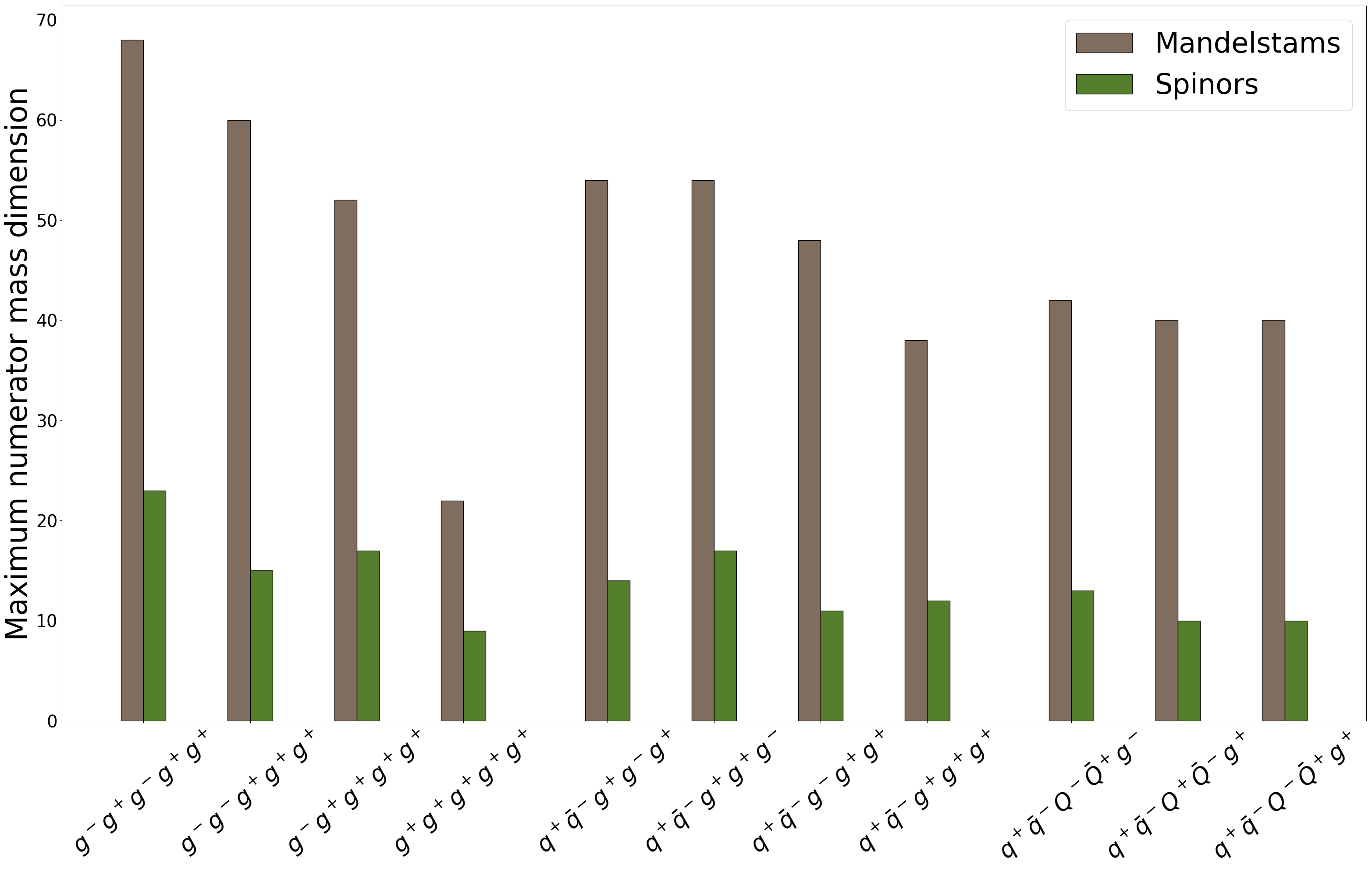}
  \caption{Maximum mass dimension of numerator ant\"aze to be reconstructed for each helicity configuration when using Mandelstam invariants or spinor variables.}
  \label{figure:maximum_numerator_mass_dim_comparison}
\end{figure}

\begin{figure}[t]
  \centering
  \includegraphics[height=67.5mm, width=120mm]{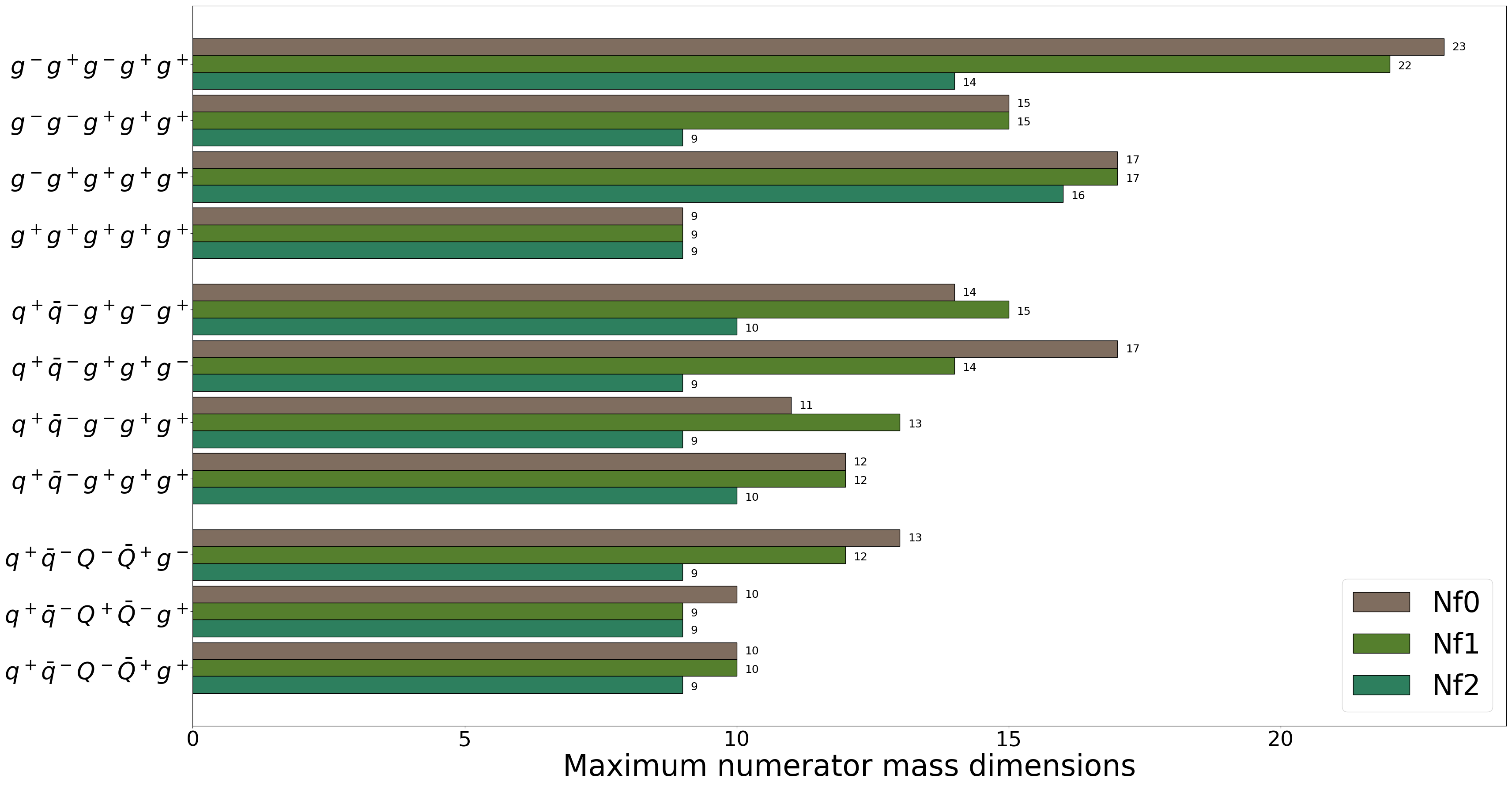}
  \caption{Maximum mass dimension of the numerator ansatze. This can be equivalently thought of as the maximum degree of the polynomials in the angle and square spinor brackets. The `Nf' labels refer to the respective terms in the expansion of Eq.~\ref{eq:nf_expansion_2l}.}
  \label{figure:maximum_numerator_degrees}
\end{figure}

Figure \ref{figure:maximum_numerator_mass_dim_comparison} shows the largest mass dimension of the numerators appearing in the rational coefficients of the various components of the two-loop five-parton finite remainders which need to be reconstructed either in terms of Mandelstam invariants via finite-field methods over a single denominator, or in terms of spinor variables via Gaussian elimination of partial-fractioned systems in specific kinematic configurations. In particular, the two largest systems to reconstruct have respectively mass dimension $d=68$ for the former case, which by Eq.~\ref{eq:ansatz_size} yields a system of size $132\,720$\footnote{The actual number of evaluations used in Ref.~\cite{Abreu:2019odu} was 94,696 instead of the predicted 132,720, because of subtleties and optimisations in their reconstruction algorithm.}, and mass dimension $d=23$ for the latter case, which gives a system of size $2\,288$\footnote{Note that Eq.~\ref{eq:ansatz_size} requires even mass dimensions. In this case the ansatz is technically for mass dimension 23 and phase weights [1, -1, 1, 1, 0].}. Since the latter case corresponds to the largest numerator in a partial-fractioned ansatz, i.e.~max($[\mathcal{N}_k]$), whereas the former to a single fraction expression, i.e.~$[\mathcal{N_{LCD}}]$, for a fair comparison we should really sum over the size of the systems for all numerators in said ansatz, i.e.~$\sum_k[\mathcal{N}_k]$. Yet, even then they only add up to about $6\,000$. Figure~\ref{figure:maximum_numerator_degrees} shows in more details the largest numerator mass dimensions for the spinor expressions. Note that common numerator factors are not taken into account, since they multiply the ansatz and are not parametrised by the latter.

A couple of further elucidations are in order. Firstly, we always solve the linear systems with an in-house implementation of a partially-pivoted row-reduction algorithm on GPGPUs, which requires double precision floats.

Secondly, the largest linear systems, such as the one of size $2\,288$, are not generated in any singular limit. These phase-space points are only required to 16 digits and are clearly relevant for all coefficients, whereas the specific phase-space points for the other terms in the partial-fraction decomposition may be required with higher precision and depend on which residues need to be fitted. Therefore, the latter type may differ from one coefficient to another depending on which poles appear. However, note that the case used above, where the number of free parameters in these numerators add up to about $4\,000$, contains 20 of the 30 poles of Eq.~\ref{eq:list_of_spinor_variables}. It is then reasonable to increase this number of $50\%$ to account for the remaining 10 poles. Still, this does not change the ball-park of up to $10\,000$ evaluations.

Thirdly, although most phase-space points are required with higher than double precision, they are also counted multiple times. For instance, if an ansatz takes the form $N_1/⟨12⟩^2+N_2/⟨12⟩$, in the sum over the number of free parameters for the numerators $N_k$ both $N_1$ and $N_2$ will appear, even though the same high-precision evaluations in the limit of small $⟨12⟩$ can be used for both coefficients. The true number of required evaluations would be max($[N_1]$, $[N_2]$) instead of $[N_1]$ + $[N_2]$. However, by performing this over-counting the estimate is closer to the number of equivalent double-precision evaluations need. In general the precision required depends linearly on how many sub-leading orders of a pole need to be resolved. In practice we used fixed precision evaluations with 300 digits and picked $\epsilon \sim 10^{-30}$ for the singular limits (Eq.~\ref{eq:singular_limits}) to avoid stability issues, but in theory $\epsilon \sim 10^{-16}$ and as few as 16 digits of precision per pole order could still allow to solve the systems with double precision. However, these type of stability considerations would need to be performed directly on the numerical code and not on already reconstructed expressions.

Overall, we find that the number of numerical evaluation required to perform the analytical reconstruction is reduced from about $100\,000$ to about $10\,000$. Considering the quoted evaluation time of 4.5 min per phase-space point in Ref.~\cite{Abreu:2019odu}, this could potentially make the calculation cheaper, depending on the additional cost of higher precision executions. Parity in this case would be obtained for an average floating-point evaluation $10\times$ slower than a finite-field one. The expectation is that as the complexity of the expressions increases, for instance as a consequence of an increased number of scales, the brute-force single-numerator reconstruction will eventually become more expensive.

\subsection{Numerical stability}

The ultimate aim of expressions such as those considered in this paper is for them to be used for predictions of physical observables, such as differential cross-sections. This requires a phase-space integration to be carried out with Monte-Carlo algorithms. In calculations of loop amplitudes the numerical accuracy of the results has been a source of concern due to inherent instabilities of the methods or expressions used.

\begin{figure}[t]
  \centering
  \includegraphics[height=47.5mm, width=120mm]{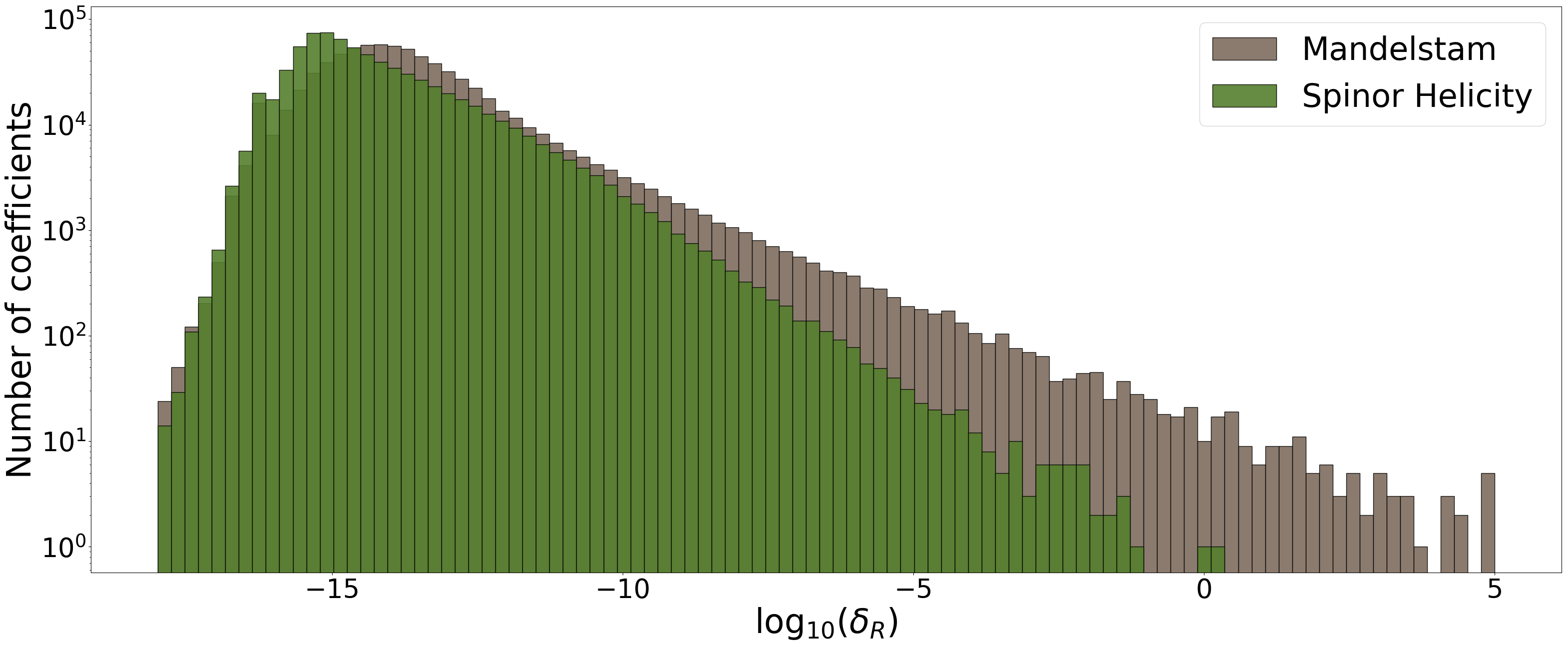}
  \caption{Relative errors over a thousand phase-space points in a collider configuration of double-precision evaluations of spinor and Mandelstam expressions.} % The x-axis is logarithmic and the error is computed w.r.t.~a high precision evaluation.
  \label{figure:relative_error}
\end{figure}

In this section, we study the double-precision floating-point numerical stability of the two-loop five-parton rational coefficients, comparing our expressions to those used as input. We do this by evaluating both sets of coefficients over $1\,000$ phase-space points. These are unweighted LO events for three-jet production at the LHC, with a center of mass energy of $14\;{\rm TeV}$, and with a jet cut of $20\;{\rm GeV}$. Care is taken that both momentum conservation and on-shell relations are satisfied to one part in $10^{16}$ for the input momenta. For each phase-space point the precision is then extended to $64$ decimal digits, the coefficients are re-evaluated and the relative error is computed as standard 
\begin{equation}
  \delta_R = \frac{(\text{double-precision evaluation}) - (\text{high-precision evaluation})}{(\text{high-precision evaluation})} \, .
\end{equation}
We also check that in the high precision evaluation the Mandelstam and spinor expressions agree at least for 16 significant digits. This is done at all $1\,000$ phase-space points for each of the $740$ coefficients for both Mandelstam and spinor representations. The resulting two sets of $740\,000$ data points are plotted in Figure~\ref{figure:relative_error}.

In the ideal case, all points would lie around the peaks at $-15$ on the x-axis, signifying little to no precision loss. However, we see that in practice this is not the case, with both curves displaying long tails extending towards the positive x-axis. Data points around 0 signify evaluations which ended up being $O(1)$ away from their true value: no significant digit is left and the output is pure numerical noise. It is apparent that the spinor expressions have an improved behaviour in the tail of the distribution.

Furthermore, the numerical stability is drastically increased for evaluations in soft or collinear regions of phase space. Evaluations in the collinear limit for 3-jet production at NNLO are rare, but will be important for 2-jet production at $\text{N}^3\text{LO}$, where these amplitudes appear as real-virtual corrections. Evaluations in the soft limit will be probed in the case of a large disparity of jet transverse momentum scales, which is relevant already for NNLO 3-jet production. For floating-point evaluations with $v_i \sim 10^{-x}$ and $v_{j≠i} \sim O(1)$ one may reasonably expect to lose $x$ digits of precision per order of the $v_i$ pole above the true one for the given expression.
% This effect is on top of the one consider until now, where precision is lost even out of any special kinematic configuration.

The presented expressions eliminate this type of instability by never having a pole of higher order than necessary. For instance, if we consider the first example in the Appendix, the Mandelstam expression \ref{eq:full_example1_mandelstam} has a spurious double pole in $s_{12}$ and a spurious triple pole in $(s_{12}+s_{23}-s_{45})\sim s_{13}$, whereas the spinor expression \ref{eq:full_example1_spinor} manifestly has no pole in the $s_{12}$ channel and has a triple pole in the spinor variable $[13]$, which for real kinematics corresponds to $s_{13}^{3/2}$.

\section{Conclusion}\label{sec:conclusion}

We presented spinor-helicity expressions for all rational coefficients of the two-loop five-parton finite remainders. These were obtained by analysing the numerical behaviour of the original expressions in Mandelstam variables \cite{Abreu:2019odu} over specific configurations of complex momentum space and by solving compact linear systems, following the strategies presented in Ref.~\cite{DeLaurentis:2019phz}.

The expressions we presented are more compact, faster to evaluate and numerically more stable, despite not being subject to any post-processing. These improvements are related to the absence of poles of higher order than necessary in the reconstructed expressions. This makes the pole structure more manifest and facilitates its physical interpretation.

Furthermore, our results required about one order of magnitude fewer numerical evaluations compared to the one originally employed, although often with higher precision. This is achieved by performing the partial-fraction decomposition before the reconstruction and not afterwards. As a result, the systems which need to be considered are drastically reduced in size. For applications to other processes in the future this reduction could prove crucial, given the time required to perform the numerical calculation in the first place.

Future applications could be the analytic reconstruction of finite remainder coefficients in two-loop processes at five-point with an external mass or at six-point.

\begin{acknowledgments}
We would like to thank Ben Page for useful discussions and Harald Ita for comments on an early draft of this paper.
\end{acknowledgments}

\pagebreak

\appendix
\section{Appendix: concrete examples}\label{sec:examples}

The following two examples are taken directly from the results provided in the accompanying files. They are labelled by particle content, helicity, term in the $N_f$ expansion (see Eq.~\ref{eq:nf_expansion_2l}), and base-1 index in the list of pentagon-function coefficients (see Eq.~\ref{eq:sum_over_coeff_pentagon_func}).

\subsection{Expression for $r_{3}$ in $\mathcal{R}^{(2)[N_f^1]}_{u^{+}\overline{u}^{-}g^{+}g^{-}g^{+}}$ }

The Mandelstam expression for the coefficient is
\begin{gather*}\label{eq:full_example1_mandelstam}
r_{3}=-1/4+s_{34}^3/(4(-s_{15}+s_{23}+s_{34})^3)- \numberthis \\
(3s_{34}^2)/(4(-s_{15}+s_{23}+s_{34})^2)+(3s_{34})/(4(-s_{15}+s_{23}+s_{34}))-\\
(3s_{15}s_{34}s_{45})/(4(-s_{15}+s_{23}+s_{34})(s_{12}+s_{23}-s_{45})^2)+\\
(3s_{15}s_{34}^2s_{45})/(4(-s_{15}+s_{23}+s_{34})^3(s_{12}+s_{23}-s_{45}))-\\
(3s_{15}s_{34}s_{45})/(4(-s_{15}+s_{23}+s_{34})^2(s_{12}+s_{23}-s_{45}))+\\
(3s_{15}^2s_{34}s_{45}^2)/(4(-s_{15}+s_{23}+s_{34})^3(s_{12}+s_{23}-s_{45})^2)+\\
(s_{15}^3s_{45}^3)/(4(-s_{15}+s_{23}+s_{34})^3(s_{12}+s_{23}-s_{45})^3)+\\
(3s_{15}s_{34}s_{45}(-s_{15}+s_{34}+s_{45}))/(4(-s_{15}+s_{23}+s_{34})^2\\
(s_{12}+s_{23}-s_{45})^2)+(3s_{15}^2s_{45}^2(-s_{15}+s_{34}+s_{45}))/\\
(4(-s_{15}+s_{23}+s_{34})^2(s_{12}+s_{23}-s_{45})^3)+\\
(3s_{15}s_{45}(s_{15}^2+s_{34}^2-3s_{15}s_{45}+s_{45}^2+2s_{34}(-s_{15}+s_{45})))/\\
(4(-s_{15}+s_{23}+s_{34})(s_{12}+s_{23}-s_{45})^3)+\\
(-s_{23}^3+3s_{23}^2s_{45}-3s_{23}s_{45}(-s_{15}+s_{45})+\\
s_{45}(3s_{15}(s_{15}-s_{34})-6s_{15}s_{45}+s_{45}^2))/\\
(4(s_{12}+s_{23}-s_{45})^3)+(-(s_{12}+s_{34})/(4s_{12}^2(-s_{15}+s_{23}+s_{34}))-\\
s_{34}^2/(4s_{12}^2(-s_{15}+s_{23}+s_{34})(s_{12}+s_{23}-s_{45}))-\\
(s_{15}^2s_{45}^2(-s_{15}+s_{34}+s_{45})^2)/(4s_{12}^2(-s_{15}+s_{23}+s_{34})^3\\
(s_{12}+s_{23}-s_{45})^3)+(s_{15}^2+s_{23}^2-s_{34}^2+s_{15}(s_{23}-6s_{45})-\\
3s_{23}s_{45}+3s_{45}^2)/(4s_{12}^2(s_{12}+s_{23}-s_{45})^2)-\\
(s_{34}^4+s_{15}^2s_{45}^2+2s_{34}^3(-s_{15}+s_{45})+4s_{15}s_{34}s_{45}\\
(-s_{15}+s_{45})+s_{34}^2(s_{15}+s_{45})^2)/(4s_{12}^2(-s_{15}+s_{23}+s_{34})^3\\
(s_{12}+s_{23}-s_{45}))+(s_{34}(-(s_{15}(s_{34}-2s_{45}))+s_{34}(s_{34}+s_{45})))/\\
(2s_{12}^2(-s_{15}+s_{23}+s_{34})^2(s_{12}+s_{23}-s_{45}))-\\
(s_{34}(s_{12}s_{34}-s_{15}(s_{34}-2s_{45})+s_{34}(s_{34}+s_{45})))/\\
(4s_{12}^2(-s_{15}+s_{23}+s_{34})^3)+(2s_{12}s_{34}+s_{15}(-s_{34}+s_{45})+\\
s_{34}(2s_{34}+s_{45}))/(4s_{12}^2(-s_{15}+s_{23}+s_{34})^2)+\\
(s_{15}s_{45}(s_{15}^2(s_{34}-s_{45})+s_{34}(s_{34}+s_{45})^2+\\
s_{15}(-2s_{34}^2-s_{34}s_{45}+s_{45}^2)))/(2s_{12}^2(s_{15}-s_{23}-s_{34})^3\\
(s_{12}+s_{23}-s_{45})^2)+(s_{15}s_{45}(s_{15}^3-(s_{34}+s_{45})^3-\\
s_{15}^2(3s_{34}+4s_{45})+s_{15}(3s_{34}^2+7s_{34}s_{45}+4s_{45}^2)))/\\
(2s_{12}^2(-s_{15}+s_{23}+s_{34})^2(s_{12}+s_{23}-s_{45})^3)+\\
(s_{15}^3-9s_{15}^2s_{45}+(2s_{34}-s_{45})(s_{34}+s_{45})^2+\\
s_{15}(-3s_{34}^2+4s_{34}s_{45}+9s_{45}^2))/(4s_{12}^2(-s_{15}+s_{23}+s_{34})\\
(s_{12}+s_{23}-s_{45})^2)+(s_{15}^4+(s_{34}+s_{45})^4-\\
2s_{15}^3(2s_{34}+5s_{45})-2s_{15}(s_{34}+s_{45})^2(2s_{34}+5s_{45})+\\
s_{15}^2(6s_{34}^2+24s_{34}s_{45}+19s_{45}^2))/(4s_{12}^2(s_{15}-s_{23}-s_{34})\\
(s_{12}+s_{23}-s_{45})^3)+(s_{15}^3(s_{34}-3s_{45})-s_{34}(s_{34}+s_{45})^3-\\
s_{15}^2(3s_{34}^2+s_{34}s_{45}-8s_{45}^2)+s_{15}(3s_{34}^3+7s_{34}^2s_{45}+\\
s_{34}s_{45}^2-3s_{45}^3))/(4s_{12}^2(-s_{15}+s_{23}+s_{34})^2\\
(s_{12}+s_{23}-s_{45})^2)+(-s_{15}^3-s_{23}^3+s_{34}^3+4s_{34}^2s_{45}+\\
6s_{34}s_{45}^2+4s_{45}^3+s_{23}^2(s_{34}+4s_{45})+\\
s_{15}^2(-s_{23}+3s_{34}+8s_{45})-s_{23}(s_{34}^2+4s_{34}s_{45}+6s_{45}^2)-\\
s_{15}(s_{23}^2+3(s_{34}+2s_{45})^2-2s_{23}(s_{34}+3s_{45})))/\\
(4s_{12}^2(s_{12}+s_{23}-s_{45})^3))\text{tr}_5 \; ,
\end{gather*}
which ha a leaf-count of 2502. The equivalent spinor expression, with a leaf-count of 32, is
\begin{equation}\label{eq:full_example1_spinor}
  r_{3}=-\frac{[32]^3 [41]^3}{2 [31]^3 [42]^3} \; .
\end{equation}

\subsection{Expression for $r_{10}$ in $\mathcal{R}^{(2)[N_f^1]}_{g^{-}g^{+}g^{-}g^{+}g^{+}}$ }

The Mandelstam expression for the coefficient is
\begin{gather*}\label{eq:full_example2_mandelstam}
r_{10} = -2/3-(13s_{15}^4)/(6(s_{12}+s_{15}-s_{34})^4)+ \numberthis \\
(25s_{15}^3)/(4(s_{12}+s_{15}-s_{34})^3)-(53s_{15}^2)/\\
(8(s_{12}+s_{15}-s_{34})^2)+(73s_{15})/(24(s_{12}+s_{15}-s_{34}))+\\
(73s_{23})/(24(s_{12}+s_{23}-s_{45}))-(s_{34}(13s_{15}+6s_{34}-6s_{45})s_{45})/\\
(6(s_{12}+s_{15}-s_{34})(s_{12}+s_{23}-s_{45})^2)-(26s_{15}^3s_{34}s_{45})/\\
(3(s_{12}+s_{15}-s_{34})^4(s_{12}+s_{23}-s_{45}))+(71s_{15}^2s_{34}s_{45})/\\
(6(s_{12}+s_{15}-s_{34})^3(s_{12}+s_{23}-s_{45}))-(13s_{15}s_{34}s_{45})/\\
(6(s_{12}+s_{15}-s_{34})^2(s_{12}+s_{23}-s_{45}))+\\
(13s_{34}s_{45})/(8(s_{12}+s_{15}-s_{34})(s_{12}+s_{23}-s_{45}))-\\
(13s_{15}^2s_{34}^2s_{45}^2)/((s_{12}+s_{15}-s_{34})^4(s_{12}+s_{23}-s_{45})^2)-\\
(111s_{34}^2s_{45}^2)/(8(s_{12}-s_{34})(s_{12}+s_{15}-s_{34})\\
(s_{12}+s_{23}-s_{45})^2)+(22s_{34}^2(s_{34}-s_{45})s_{45}^2)/\\
((s_{12}-s_{34})(s_{12}+s_{15}-s_{34})(s_{12}+s_{23}-s_{45})^3)-\\
(26s_{15}s_{34}^3s_{45}^3)/(3(s_{12}+s_{15}-s_{34})^4(s_{12}+s_{23}-s_{45})^3)-\\
(19s_{34}^3s_{45}^3)/(2(s_{12}-s_{34})(s_{12}+s_{15}-s_{34})^2\\
(s_{12}+s_{23}-s_{45})^3)+(5s_{34}^3s_{45}^3)/(8(s_{12}-s_{34})^2\\
(s_{12}+s_{15}-s_{34})(s_{12}+s_{23}-s_{45})^3)+(7s_{34}^3(s_{34}-s_{45})s_{45}^3)/\\
((s_{12}-s_{34})(s_{12}+s_{15}-s_{34})^2(s_{12}+s_{23}-s_{45})^4)+\\
(s_{34}^3(s_{34}-s_{45})s_{45}^3)/(2(s_{12}-s_{34})^2(s_{12}+s_{15}-s_{34})\\
(s_{12}+s_{23}-s_{45})^4)-(13s_{34}^4s_{45}^4)/(6(s_{12}+s_{15}-s_{34})^4\\
(s_{12}+s_{23}-s_{45})^4)-(7s_{34}^4s_{45}^4)/(4(s_{12}-s_{34})\\
(s_{12}+s_{15}-s_{34})^3(s_{12}+s_{23}-s_{45})^4)-\\
(s_{34}^4s_{45}^4)/(8(s_{12}-s_{34})^2(s_{12}+s_{15}-s_{34})^2\\
(s_{12}+s_{23}-s_{45})^4)-(s_{34}^4s_{45}^4)/(2(s_{12}-s_{34})^3\\
(s_{12}+s_{15}-s_{34})(s_{12}+s_{23}-s_{45})^4)-\\
(26s_{34}^3s_{45}^3(s_{15}-s_{34}+s_{45}))/(3(s_{12}+s_{15}-s_{34})^3\\
(s_{12}+s_{23}-s_{45})^4)+(s_{15}s_{34}s_{45}(-3s_{15}s_{34}+3s_{15}s_{45}+\\
5s_{34}s_{45}))/(2(-s_{12}+s_{34})^3(s_{12}+s_{23}-s_{45})^2)-\\
(3(-2s_{15}s_{34}+2s_{15}s_{45}+11s_{34}s_{45}))/(8(s_{12}-s_{34})\\
(s_{12}+s_{23}-s_{45}))+(s_{34}s_{45}(71s_{15}^2-19s_{15}s_{34}+19s_{15}s_{45}+\\
48s_{34}s_{45}))/(6(s_{12}+s_{15}-s_{34})^2(s_{12}+s_{23}-s_{45})^2)+\\
(36s_{15}s_{23}-195s_{23}^2-18s_{15}s_{34}+18s_{15}s_{45}+133s_{34}s_{45})/\\
(24(s_{12}+s_{23}-s_{45})^2)+(s_{34}s_{45}(71s_{15}^2-38s_{15}s_{34}-\\
33s_{34}^2+38s_{15}s_{45}+189s_{34}s_{45}-33s_{45}^2))/\\
(6(s_{12}+s_{15}-s_{34})(s_{12}+s_{23}-s_{45})^3)-\\
(s_{34}^2s_{45}^2(21s_{34}^2-43s_{34}s_{45}+21s_{45}^2))/\\
(2(s_{12}-s_{34})(s_{12}+s_{15}-s_{34})(s_{12}+s_{23}-s_{45})^4)-\\
((-(s_{15}s_{34})+s_{15}s_{45}+5s_{34}s_{45})(s_{15}s_{45}+s_{34}(-s_{15}+s_{45}))^2)/\\
(2(s_{12}-s_{34})^3(s_{12}+s_{23}-s_{45})^3)+\\
(2s_{15}s_{34}s_{45}(-13s_{15}(s_{15}+s_{45})+s_{34}(13s_{15}+10s_{45})))/\\
(3(s_{12}+s_{15}-s_{34})^3(s_{12}+s_{23}-s_{45})^2)+\\
(s_{34}^2s_{45}^2(-104s_{15}(s_{15}+s_{45})+s_{34}(104s_{15}+17s_{45})))/\\
(6(s_{12}+s_{15}-s_{34})^3(s_{12}+s_{23}-s_{45})^3)-\\
(s_{34}^2s_{45}^2(39s_{34}^2+39(s_{15}+s_{45})^2-s_{34}(78s_{15}+125s_{45})))/\\
(3(s_{12}+s_{15}-s_{34})^2(s_{12}+s_{23}-s_{45})^4)+\\
(4s_{15}(3s_{15}-7s_{45})s_{45}+28s_{34}^2(-s_{15}+4s_{45})+\\
s_{34}(-12s_{15}^2+85s_{15}s_{45}-112s_{45}^2))/(8(s_{12}-s_{34})\\
(s_{12}+s_{23}-s_{45})^2)+(s_{15}s_{34}(24s_{15}-29s_{45})s_{45}-\\
6s_{15}^2s_{45}^2+s_{34}^2(-6s_{15}^2+29s_{15}s_{45}-5s_{45}^2))/\\
(8(s_{12}-s_{34})^2(s_{12}+s_{23}-s_{45})^2)-\\
(s_{34}s_{45}(52s_{15}(s_{15}+s_{45})^2+s_{34}^2(52s_{15}+9s_{45})-\\
s_{34}(104s_{15}^2+217s_{15}s_{45}+9s_{45}^2)))/(6(s_{12}+s_{15}-s_{34})^2\\
(s_{12}+s_{23}-s_{45})^3)+(s_{34}s_{45}(52s_{34}^3-52(s_{15}+s_{45})^3-\\
6s_{34}^2(26s_{15}+73s_{45})+3s_{34}(52s_{15}^2+177s_{15}s_{45}+\\
146s_{45}^2)))/(6(s_{12}+s_{15}-s_{34})(s_{12}+s_{23}-s_{45})^4)+\\
(-102s_{15}s_{23}^2+231s_{23}^3+s_{23}s_{34}(18s_{15}-257s_{45})+\\
s_{34}^2(66s_{15}-179s_{45})+3s_{15}s_{23}(7s_{15}-6s_{45})+\\
6s_{15}s_{45}(-6s_{15}+11s_{45})+s_{34}(36s_{15}^2-503s_{15}s_{45}+\\
179s_{45}^2))/(24(s_{12}+s_{23}-s_{45})^3)+\\
(s_{34}^3(50s_{15}-161s_{45})+s_{15}s_{45}(-12s_{15}^2+57s_{15}s_{45}-\\
50s_{45}^2)+s_{34}^2(57s_{15}^2-280s_{15}s_{45}+363s_{45}^2)+\\
s_{34}(12s_{15}^3-142s_{15}^2s_{45}+280s_{15}s_{45}^2-161s_{45}^3))/\\
(8(s_{12}-s_{34})(s_{12}+s_{23}-s_{45})^3)+\\
(s_{15}^2(12s_{15}-19s_{45})s_{45}^2+s_{15}s_{34}s_{45}(-24s_{15}^2+113s_{15}s_{45}-\\
65s_{45}^2)+s_{34}^3(19s_{15}^2-65s_{15}s_{45}+41s_{45}^2)+\\
s_{34}^2(12s_{15}^3-113s_{15}^2s_{45}+174s_{15}s_{45}^2-41s_{45}^3))/\\
(8(s_{12}-s_{34})^2(s_{12}+s_{23}-s_{45})^3)+\\
((s_{34}-s_{45})(s_{15}^3s_{45}^3+s_{15}^2s_{34}s_{45}^2(-3s_{15}+4s_{45})+\\
s_{15}s_{34}^2s_{45}(3s_{15}^2-8s_{15}s_{45}+6s_{45}^2)+\\
s_{34}^3(-s_{15}^3+4s_{15}^2s_{45}-6s_{15}s_{45}^2+4s_{45}^3)))/\\
(2(s_{12}-s_{34})^3(s_{12}+s_{23}-s_{45})^4)+\\
(s_{15}^2(16s_{15}-13s_{45})s_{45}^3+s_{34}^4(-13s_{15}^2+36s_{15}s_{45}-\\
30s_{45}^2)-4s_{15}s_{34}s_{45}^2(12s_{15}^2-25s_{15}s_{45}+9s_{45}^2)-\\
6s_{34}^2s_{45}(-8s_{15}^3+29s_{15}^2s_{45}-26s_{15}s_{45}^2+5s_{45}^3)+\\
4s_{34}^3(-4s_{15}^3+25s_{15}^2s_{45}-39s_{15}s_{45}^2+19s_{45}^3))/\\
(8(s_{12}-s_{34})^2(s_{12}+s_{23}-s_{45})^4)+\\
(s_{34}^4(-7s_{15}+19s_{45})+s_{34}^3(-13s_{15}^2+55s_{15}s_{45}-72s_{45}^2)+\\
s_{15}s_{45}^2(-6s_{15}^2+13s_{15}s_{45}-7s_{45}^2)+\\
s_{34}s_{45}(12s_{15}^3-51s_{15}^2s_{45}+55s_{15}s_{45}^2-19s_{45}^3)+\\
3s_{34}^2(-2s_{15}^3+17s_{15}^2s_{45}-34s_{15}s_{45}^2+24s_{45}^3))/\\
(2(s_{12}-s_{34})(s_{12}+s_{23}-s_{45})^4)+\\
(84s_{15}s_{23}^3-109s_{23}^4+s_{34}^3(-84s_{15}+436s_{45})+\\
s_{23}^2(-39s_{15}^2-84s_{15}s_{34}+84s_{15}s_{45}+436s_{34}s_{45})+\\
s_{34}^2(-117s_{15}^2+52s_{15}s_{45}-1526s_{45}^2)+\\
3s_{15}s_{45}(12s_{15}^2-39s_{15}s_{45}+28s_{45}^2)+\\
s_{34}(-36s_{15}^3+490s_{15}^2s_{45}-52s_{15}s_{45}^2+436s_{45}^3)+\\
s_{23}(s_{34}^2(84s_{15}-436s_{45})+6s_{15}(2s_{15}^2-13s_{15}s_{45}+\\
14s_{45}^2)+s_{34}(78s_{15}^2-68s_{15}s_{45}+436s_{45}^2)))/\\
(24(s_{12}+s_{23}-s_{45})^4)+(-73/(24s_{12}^2)+(13s_{15}^3(s_{15}-s_{34})^2)/\\
(6s_{12}^2(s_{12}+s_{15}-s_{34})^4(s_{12}+s_{23}-s_{45}))+\\
s_{34}^2/(2s_{12}^2(s_{12}-s_{34})(s_{12}+s_{23}-s_{45}))+\\
(-116s_{15}^2+73s_{15}s_{34}-6s_{34}^2)/(12s_{12}^2(s_{12}+s_{15}-s_{34})\\
(s_{12}+s_{23}-s_{45}))-(s_{15}^2(101s_{15}^2-150s_{15}s_{34}+49s_{34}^2))/\\
(12s_{12}^2(s_{12}+s_{15}-s_{34})^3(s_{12}+s_{23}-s_{45}))+\\
(s_{15}(309s_{15}^2-318s_{15}s_{34}+61s_{34}^2))/\\
(24s_{12}^2(s_{12}+s_{15}-s_{34})^2(s_{12}+s_{23}-s_{45}))+\\
(13s_{15}^2(s_{15}-s_{34})^2s_{34}s_{45})/(2s_{12}^2(s_{12}+s_{15}-s_{34})^4\\
(s_{12}+s_{23}-s_{45})^2)+(s_{15}s_{34}^3s_{45})/(2s_{12}^2(s_{12}-s_{34})^3\\
(s_{12}+s_{23}-s_{45})^2)+(9s_{34}^3s_{45})/(2s_{12}^2(s_{12}-s_{34})\\
(s_{12}+s_{15}-s_{34})(s_{12}+s_{23}-s_{45})^2)+\\
(13s_{15}(s_{15}-s_{34})^2s_{34}^2s_{45}^2)/(2s_{12}^2(s_{12}+s_{15}-s_{34})^4\\
(s_{12}+s_{23}-s_{45})^3)+(49s_{34}^4s_{45}^2)/(8s_{12}^2(s_{12}-s_{34})\\
(s_{12}+s_{15}-s_{34})^2(s_{12}+s_{23}-s_{45})^3)-(3s_{34}^4s_{45}^2)/\\
(8s_{12}^2(s_{12}-s_{34})^2(s_{12}+s_{15}-s_{34})(s_{12}+s_{23}-s_{45})^3)+\\
(13(s_{15}-s_{34})^2s_{34}^3s_{45}^3)/(6s_{12}^2(s_{12}+s_{15}-s_{34})^4\\
(s_{12}+s_{23}-s_{45})^4)+(7s_{34}^5s_{45}^3)/(4s_{12}^2(s_{12}-s_{34})\\
(s_{12}+s_{15}-s_{34})^3(s_{12}+s_{23}-s_{45})^4)+\\
(s_{34}^5s_{45}^3)/(8s_{12}^2(s_{12}-s_{34})^2(s_{12}+s_{15}-s_{34})^2\\
(s_{12}+s_{23}-s_{45})^4)+(s_{34}^5s_{45}^3)/(2s_{12}^2(s_{12}-s_{34})^3\\
(s_{12}+s_{15}-s_{34})(s_{12}+s_{23}-s_{45})^4)+\\
(3s_{34}^3s_{45}(-12s_{34}+11s_{45}))/(4s_{12}^2(s_{12}-s_{34})\\
(s_{12}+s_{15}-s_{34})(s_{12}+s_{23}-s_{45})^3)+\\
(s_{34}^4s_{45}^2(-3s_{34}+11s_{45}))/(8s_{12}^2(s_{12}-s_{34})^2\\
(s_{12}+s_{15}-s_{34})(s_{12}+s_{23}-s_{45})^4)+\\
(s_{34}^4s_{45}^2(-21s_{34}+22s_{45}))/(4s_{12}^2(s_{12}-s_{34})\\
(s_{12}+s_{15}-s_{34})^2(s_{12}+s_{23}-s_{45})^4)+\\
(s_{34}^2(-2s_{15}(s_{34}-5s_{45})+s_{34}s_{45}))/(8s_{12}^2(s_{12}-s_{34})^2\\
(s_{12}+s_{23}-s_{45})^2)+(s_{34}(-6s_{15}s_{34}-15s_{34}^2+8s_{15}s_{45}+\\
17s_{34}s_{45}))/(8s_{12}^2(s_{12}-s_{34})(s_{12}+s_{23}-s_{45})^2)+\\
(s_{34}^3s_{45}(42s_{34}^2-93s_{34}s_{45}+52s_{45}^2))/(8s_{12}^2(s_{12}-s_{34})\\
(s_{12}+s_{15}-s_{34})(s_{12}+s_{23}-s_{45})^4)-\\
(s_{34}^2(s_{34}-s_{45})(s_{15}^2(s_{34}-s_{45})^2+3s_{34}^2s_{45}^2+\\
3s_{15}s_{34}s_{45}(-s_{34}+s_{45})))/(2s_{12}^2(s_{12}-s_{34})^3\\
(s_{12}+s_{23}-s_{45})^4)+(s_{34}^2(s_{15}^2(s_{34}-s_{45})^2+\\
3s_{34}^2s_{45}^2+4s_{15}s_{34}s_{45}(-s_{34}+s_{45})))/\\
(2s_{12}^2(s_{12}-s_{34})^3(s_{12}+s_{23}-s_{45})^3)+\\
(232s_{15}+268s_{23}-73(s_{34}+s_{45}))/(24s_{12}^2(s_{12}+s_{23}-s_{45}))-\\
(13s_{34}(-s_{15}+s_{34})s_{45}(8s_{15}^2(s_{15}+s_{45})+\\
s_{34}^2(8s_{15}+3s_{45})-s_{15}s_{34}(16s_{15}+23s_{45})))/\\
(12s_{12}^2(s_{12}+s_{15}-s_{34})^3(s_{12}+s_{23}-s_{45})^3)+\\
(309s_{15}^3+s_{15}s_{34}(183s_{34}-262s_{45})-159s_{15}^2(3s_{34}-s_{45})+\\
3s_{34}^2(-5s_{34}+43s_{45}))/(24s_{12}^2(s_{12}+s_{15}-s_{34})\\
(s_{12}+s_{23}-s_{45})^2)+(-309s_{15}^2-426s_{23}^2-\\
6s_{15}(48s_{23}-53s_{34})+195s_{23}(s_{34}+s_{45})+\\
s_{34}(-73s_{34}+195s_{45}))/(24s_{12}^2(s_{12}+s_{23}-s_{45})^2)+\\
(s_{34}^2s_{45}^2(-78s_{34}^3+78s_{15}^2(s_{15}+s_{45})+\\
2s_{34}^2(117s_{15}+86s_{45})-s_{15}s_{34}(234s_{15}+229s_{45})))/\\
(12s_{12}^2(s_{12}+s_{15}-s_{34})^3(s_{12}+s_{23}-s_{45})^4)-\\
(s_{15}(-s_{15}+s_{34})(26s_{15}^2(s_{15}+s_{45})+s_{34}^2(26s_{15}+67s_{45})-\\
s_{15}s_{34}(52s_{15}+249s_{45})))/(12s_{12}^2(s_{12}+s_{15}-s_{34})^3\\
(s_{12}+s_{23}-s_{45})^2)+(s_{34}^3(46s_{15}-51s_{45})-\\
2s_{15}^3(101s_{15}+75s_{45})-2s_{15}s_{34}^2(147s_{15}+106s_{45})+\\
s_{15}^2s_{34}(450s_{15}+569s_{45}))/(24s_{12}^2(s_{12}+s_{15}-s_{34})^2\\
(s_{12}+s_{23}-s_{45})^2)+(2s_{15}^2(6s_{34}^2-6s_{34}s_{45}+s_{45}^2)+\\
s_{15}s_{34}(22s_{34}^2-57s_{34}s_{45}+19s_{45}^2)+\\
s_{34}^2(15s_{34}^2-54s_{34}s_{45}+37s_{45}^2))/(4s_{12}^2(s_{12}-s_{34})\\
(s_{12}+s_{23}-s_{45})^3)+(s_{34}(8s_{34}^2s_{45}(-2s_{34}+5s_{45})+\\
8s_{15}^2(2s_{34}^2-3s_{34}s_{45}+s_{45}^2)+\\
s_{15}s_{34}(11s_{34}^2-70s_{34}s_{45}+43s_{45}^2)))/\\
(8s_{12}^2(s_{12}-s_{34})^2(s_{12}+s_{23}-s_{45})^3)+\\
(-101s_{15}^4+150s_{15}^3(2s_{34}-s_{45})+\\
s_{15}^2(-294s_{34}^2+606s_{34}s_{45}-49s_{45}^2)+\\
2s_{15}s_{34}(46s_{34}^2-171s_{34}s_{45}+57s_{45}^2)+\\
3s_{34}^2(s_{34}^2-74s_{34}s_{45}+59s_{45}^2))/\\
(12s_{12}^2(s_{12}+s_{15}-s_{34})(s_{12}+s_{23}-s_{45})^3)+\\
(-19s_{34}^5+4s_{15}^2s_{45}^3+6s_{15}s_{34}s_{45}^2(-6s_{15}+5s_{45})+\\
9s_{34}^4(-5s_{15}+13s_{45})+s_{34}^3(-40s_{15}^2+180s_{15}s_{45}-\\
183s_{45}^2)+s_{34}^2s_{45}(72s_{15}^2-153s_{15}s_{45}+61s_{45}^2))/\\
(8s_{12}^2(s_{12}-s_{34})(s_{12}+s_{23}-s_{45})^4)+\\
(52s_{15}^3(s_{15}+s_{45})^2+s_{34}^4(52s_{15}+72s_{45})-\\
16s_{15}^2s_{34}(13s_{15}^2+63s_{15}s_{45}+37s_{45}^2)+\\
s_{34}^3(-208s_{15}^2-944s_{15}s_{45}+66s_{45}^2)+\\
3s_{15}s_{34}^2(104s_{15}^2+592s_{15}s_{45}+259s_{45}^2))/\\
(24s_{12}^2(s_{12}+s_{15}-s_{34})^2(s_{12}+s_{23}-s_{45})^3)+\\
(s_{34}s_{45}(156s_{34}^4+156s_{15}^2(s_{15}+s_{45})^2-\\
6s_{34}^3(104s_{15}+193s_{45})-6s_{15}s_{34}(104s_{15}^2+255s_{15}s_{45}+\\
125s_{45}^2)+s_{34}^2(936s_{15}^2+2250s_{15}s_{45}+817s_{45}^2)))/\\
(24s_{12}^2(s_{12}+s_{15}-s_{34})^2(s_{12}+s_{23}-s_{45})^4)+\\
(-52s_{34}^5+52s_{15}^2(s_{15}+s_{45})^3+4s_{34}^4(65s_{15}+321s_{45})-\\
s_{34}^3(520s_{15}^2+3252s_{15}s_{45}+2721s_{45}^2)-\\
2s_{15}s_{34}(130s_{15}^3+687s_{15}^2s_{45}+828s_{15}s_{45}^2+271s_{45}^3)+\\
s_{34}^2(520s_{15}^3+3312s_{15}^2s_{45}+3825s_{15}s_{45}^2+907s_{45}^3))/\\
(24s_{12}^2(s_{12}+s_{15}-s_{34})(s_{12}+s_{23}-s_{45})^4)+\\
(s_{34}(-4s_{15}^2(5s_{34}-2s_{45})(s_{34}-s_{45})^2+3s_{34}^2s_{45}\\
(5s_{34}^2-22s_{34}s_{45}+13s_{45}^2)+\\
s_{15}(-9s_{34}^4+75s_{34}^3s_{45}-99s_{34}^2s_{45}^2+33s_{34}s_{45}^3)))/\\
(8s_{12}^2(s_{12}-s_{34})^2(s_{12}+s_{23}-s_{45})^4)+\\
(202s_{15}^3+340s_{23}^3+2s_{15}(77s_{23}^2-129s_{23}s_{34}+\\
s_{34}(156s_{34}-397s_{45}))+2s_{23}s_{34}(61s_{34}-231s_{45})-\\
231s_{23}^2(s_{34}+s_{45})+s_{34}^2(-13s_{34}+283s_{45})+\\
2s_{15}^2(107s_{23}+75(-3s_{34}+s_{45})))/(24s_{12}^2\\
(s_{12}+s_{23}-s_{45})^3)-(52s_{15}^4+52s_{15}^3(s_{23}-4s_{34}+2s_{45})+\\
2s_{15}^2(32s_{23}^2+192s_{34}^2-489s_{34}s_{45}+32s_{45}^2+\\
32s_{23}(-3s_{34}+s_{45}))+109(s_{23}^4-s_{23}s_{34}^2(s_{34}-7s_{45})+\\
s_{23}^2s_{34}(s_{34}-3s_{45})-s_{23}^3(s_{34}+s_{45})+\\
s_{34}^2(s_{34}^2-11s_{34}s_{45}+10s_{45}^2))+\\
s_{15}(25s_{23}^3-50s_{23}^2s_{34}+s_{23}s_{34}(75s_{34}-452s_{45})-\\
4s_{34}(25s_{34}^2-339s_{34}s_{45}+113s_{45}^2)))/\\
(24s_{12}^2(s_{12}+s_{23}-s_{45})^4))\text{tr}_5 \, ,
\end{gather*}
which has a leaf-count of 6848. The equivalent spinor expression, with a leaf-count of 274, is
\begin{gather*}\label{eq:full_example2_spinor}
r_{10}=-1\frac{[12]^3[15][23]⟨25⟩^3[35]^3}{[13]^4[25]⟨5|1+2|5]^3}+\frac{97}{12}\frac{[12]^4⟨25⟩[35]^4}{[13]^4[25]^3⟨5|1+2|5]}+ \numberthis \\
\frac{13}{3}\frac{[12]^4⟨15⟩[15][35]^4}{[13]^4[25]^4⟨5|1+2|5]}+\frac{1}{4}\frac{[12]^4⟨15⟩[15]⟨25⟩[35]^4}{[13]^4[25]^3⟨5|1+2|5]^2}+\\
-\frac{3}{2}\frac{[12]^2⟨25⟩^2[25][35]^2}{[13]^2[25]⟨5|1+2|5]^2}+\frac{7}{4}\frac{[12]^3⟨25⟩^2[35]^3}{[13]^3[25]⟨5|1+2|5]^2}+\\
-\frac{43}{3}\frac{[12]^3⟨25⟩[35]^3}{[13]^3[25]^2⟨5|1+2|5]}-\frac{25}{3}\frac{[12]^3⟨15⟩[15][35]^3}{[13]^3[25]^3⟨5|1+2|5]}+\\
-\frac{3}{2}\frac{[12]⟨25⟩[25][35]}{[13][25]⟨5|1+2|5]}+4\frac{[12]^2⟨25⟩[35]^2}{[13]^2[25]⟨5|1+2|5]}+\\
-\frac{15}{2}\frac{[12]^2[35]^2}{[13]^2[25]^2}+\frac{7}{2}\frac{[12][35]}{[13][25]}-\frac{2}{3} \, .
\end{gather*}

\bibliography{article}
\bibliographystyle{JHEP}

\end{document}